\documentclass[doublecol]{epl2} 
\usepackage{color}
\usepackage{graphicx}

\usepackage{url}

\usepackage{xcolor}
\newcommand{\jm}[1]{\textcolor{purple}{#1}}

\title{20 Years of Ordinal Patterns: Perspectives and Challenges}
\shorttitle{20 years of ordinal patterns} 

\author{ 
Inmaculada Leyva  \inst{1,2}  \and
Johann H. Mart\'inez   \inst{3} \and
Cristina Masoller \inst{4} \and 
Osvaldo A. Rosso  \inst{5} \and \\
Massimiliano Zanin \inst{3}.}

\institute{                    
  \inst{1}  Universidad Rey Juan Carlos, Calle Tulip\'an s/n, 28933 M\'ostoles, Madrid, Spain \\
\inst{2} Center for Biomedical Technology, Universidad Polit\'ecnica de Madrid, 28223 Pozuelo de Alarc\'on, Madrid, Spain \\
  \inst{3} Instituto de F\'isica Interdisciplinar y Sistemas Complejos IFISC (CSIC-UIB), Campus UIB, 07122 Palma de Mallorca, Spain\\
  \inst{4} Department of Physics, Universitat Politecnica de Catalunya, Rambla St. Nebridi 22, Terrassa 08222, Spain\\
  \inst{5} Instituto de F\'{\i}sica, Universidade Federal de Alagoas (UFAL), Maceió, Alagoas 57072-970, Brasil.
}

\shortauthor{I. Leyva, J. Martinez, C. Masoller, O. A. Rosso, M. Zanin}

\abstract{
In 2002, in a seminal article, Christoph Bandt and Bernd Pompe proposed a new methodology for the analysis of complex time series, now known as Ordinal Analysis. The ordinal methodology is based on the computation of symbols (known as ordinal patters) which are defined in terms of the temporal ordering of data points in a time series, and whose probabilities are known as ordinal probabilities. With the ordinal probabilities the Shannon entropy can be calculated, which is the permutation entropy. Since it was proposed, the ordinal method has found applications in fields as diverse as biomedicine and climatology. However, some properties of ordinal probabilities are still not fully understood, and how to combine the ordinal approach of feature extraction with machine learning techniques for model identification, time series classification or forecasting, remains a challenge. The objective of this perspective article is to present some recent advances and to discuss some open problems.
}

\begin{document}

\maketitle

\section{Introduction}
Since it was published in 2002, the seminal paper by Christoph Bandt and Bernd Pompe~\cite{bp_prl_2002} has been cited more than 17000 times, and the number of citations per year continues to grow, nearly exponentially. The aim of this article is to present a brief overview of the ordinal methodology and to present some examples of applications in different fields. 

An important challenge in time series analysis is how can we determine if a time series was generated by a low-dimensional, possibly chaotic dynamics, or by a high-dimensional, possibly stochastic dynamics.

Stochastic and chaotic time series display some characteristic which make them almost indistinguishable:
{\it a)} A wide-band power spectrum;
{\it b)} A delta like auto-correlation function;
{\it c)} An irregular behavior of the measured signals.
Chaotic systems display {\it “sensitivity to initial conditions”} which manifests instability 
in the phase space and leads to non-periodic motion.
They display long-term unpredictability despite the deterministic character of the temporal trajectory.
In a system undergoing chaotic motion, two trajectories starting in two neighboring points in phase space will diverge.
Let $x_1(t)$ and $x_2(t)$ be two such points, located within a ball of radius ${\mathcal R}$ at time $t$. Further, let us assume that these two points cannot be resolved within the ball due to the resolution of the measurement instrument.
At a later time, $t^{\star}$, the distance between the points will 
grow (as the points diverge due to chaotic motion),
and at time $t^{\star}$ the points can become experimentally distinguishable. 
This implies that chaotic motion reveals some   
information 
that was not available at earlier times.
Then, we can think of chaos as an {\it information source}, and we can use quantifiers based on Information Theory to characterize and better understand chaotic systems. This approach, combined with the use of symbolic analysis, was first explored by Bandt and Pompe (BP) in 2002~\cite{bp_prl_2002}. Since its success with a well-known chaotic system, the Logistic map, the BP approach has been used to analyze a wide range of complex data sets. It would not possible be to review the many applications that the BP methodology has found in the last 20 years, so here we limit to discuss a few examples.

The organization of this perspective article is as follows: first, we discuss two information-theory quantifiers, the entropy and the complexity, and introduce the concept of ordinal patterns. Then, we discuss applications of these concepts in different fields. We finalize with a discussion of future promising lines of research.

\section{Entropy, complexity and ordinal patterns}

A well-known quantifier is normalized entropy, $H[P] = S[P] / S_{max}$, with $S$ as the Shannon's entropy~\cite{Shannon_1948}, and $S_{max}$ the system's maximal entropy
that is zero when the information (knowledge) about the underlying process described by the probability distribution function (PDF) $P$ is maximal and the outcome of a measure can be predicted with complete certainty. 
In contrast, when physical process follows a uniform distribution, $P_e$, the knowledge is minimal and the entropy is maximum.
A related concept is the
Statistical Complexity, $C[P]$, defined as the product of the entropy, $H[P]$, with the {{disequilibrium}}, $Q[P]$, which is the distance between the distribution $P$ and the equilibrium distribution, $P_e$. The distance between $P$ and $P_e$ is often defined in terms of the Jensen-Shannon divergence \cite{MPR_PhysicaA_2004}.

In physics, complexity notion starts by considering a perfect crystal and a isolated gas as two examples of simple states of matter and therefore, as systems with zero complexity. The structure of a perfect crystal is completely described by minimal information (i.e., distances and symmetries that define the elementary cell) and the probability distribution of the accessible states is represented by a delta function (the state of perfect symmetry has probability equal one, while any other state has zero probability). In contrast, the probability distribution of the possible states of an ideal gas in equilibrium is the "simple" uniform distribution. 
Therefore, both situations have minimum complexity, and then it is clear that a suitable measure of complexity can not be made only in terms of ``disorder", nor only in terms of ``information". However, the definition of statistical complexity as the product of $S$ and $Q$ satisfies the condition of being zero in both limits (for an ideal crystal, $C=0$ because $H=0$, while for an ideal gas in equilibrium, $C=0$ because $Q=0$). 

The statistical complexity $C=H \times Q$ quantifies the existence of non-trivial structures.  
In the cases of perfect order and total randomness $C[P]=0$ means the data possess no structure. In between these two extreme instances, a large range of possible values quantifies the level of structure in the data.
The statistical complexity is able to detect subtle 
details of the dynamical processes that generate the data~\cite{Complex_PRL_2007}.  
This is due to the fact that for a given value of the entropy, there is a range of possible values of complexity.

It is interesting to note that \jm{$C$} gives additional information in relation to the entropy, due to its dependence on two distributions (the one associated with the system under analysis, $P$, and the equilibrium distribution, $P_e$). An important property of $C[P]$ is that its value does not change with different ordering of the PDF. Moreover, the extreme complexity values, $C_{\mathrm{min}}$ and $C_{\mathrm{max}}$, depend only on the dimension of the probability space and on the functional forms adopted for {{$H$ and $Q$}}~\cite{Cotas-PhysicaA_2006}. Then, one can evaluate the so-called “Entropy-Complexity Plane, $H \times C$” in order to represent the system's dynamics in this plane. 

There is no unique answer for the best procedure to associate a PDF to a time series. However, when we consider a sequence of measurements taken in time causal order, the temporal structure of the time-series needs to be preserved in the associated PDF. For this reason, the procedure proposed by Bandt and Pompe (BP)~\cite{bp_prl_2002} is very popular, because it allows to associate a time-causal PDF to the time series.
In BP approach, from the time series a sequence of symbols known as ``ordinal patterns'' (OPs) is calculated, and the BP probability distribution is obtained by the frequency histogram of symbols. The OPs are defined by the sequential order of the data points in the time series, without considering the actual values of the data points (see \cite{bp_prl_2002} for details).

The OPs represent a natural alphabet for a time series~\cite{amigo_book}, and are defined in terms of two parameters, the ``length'' of the pattern, $D$, and the lag, $\tau$, between the data points in the time series that are used to define the patterns. The number of possible ordinal patterns (i.e., ``letters'' in the alphabet) grows as $D$!. 

An important property of the BP--PDF is that it is invariant under monotonic transformations, and also, it incorporates time causality in a natural way.
Moreover, the only condition for applicability of the BP methodology is a very weak stationary assumption \cite{bp_prl_2002}.
When the BP--PDF is used to compute $H$ and $C$, they are denoted as ``permutation entropy'' and ``causal complexity'', respectively, which span the so-called ``time causal Entropy-Complexity plane",  $H \times C$.

The $H \times C$ plane has been shown to be a very useful diagnostic tool to discriminate between chaotic and stochastic time series~\cite{Complex_PRL_2007}, since the $H$ and $C$ quantifiers have distinctive behaviors for different types of dynamics.
Chaotic maps have intermediate permutation entropy $H$ while the complexity $C$ is close to the maximum value, $C_{\mathrm{max}}$  (see Fig. 1 in~\cite{Complex_PRL_2007}).
Similar behavior is still observed when the time series is contaminated with small or moderate amount of uncorrelated or correlated noise (see 
~\cite{HxC_Noise_EPJB_2012}). 
In contrast, fully uncorrelated stochastic time series are located in the $H \times C$ plane quite close to extreme point $(H,C) \approx (1,0)$. Pure correlated stochastic time series present decreasing values of entropy $H$ with increasing correlation strength, and associate increase of the complexity $C$ value at intermediate value between $C_{\mathrm{min}}$ and $C_{\mathrm{max}}$ (see~\cite{Complex_PRL_2007, HxC_Noise_EPJB_2012}).

The ordinal pattern probabilities and the $H \times C$ plane have been used in many different applications, among which we can mention the analysis and characterization of stochastic and coherence resonances~\cite{Resonancia_EPJB_2009,Resonancia_PRE_2009};
of the nonlinear dynamics of chaotic lasers~\cite{Tiana_2010,Soriano_2010,Kane_2014};
of the Libor market~\cite{Libor_EPJB_2015};
of large-scale atmospheric patterns~\cite{barreiro_chaos_2011,deza_chaos_2015};
of handwriten signatures~\cite{Signature-PlosOne_2016};
of the Ecosystem Gross Primary productivity~\cite{Holger-PlosOne_2016};
of transportation media~\cite{Vehiculos_Ad_HocNetwork_2019};
of electric load consumption ~\cite{Electric_PhysicaA_2017};
of cardiac~\cite{parlitz_2012,amigo_2013} and brain signals {{(Electroencephalography (EEG) and Magnetoencephalography (MEG)~\cite{EEG_Chaos_2018,MEG_Chaos_2020,parlitz_2021})}}; of textures in synthetic aperture radar (SAR) imagery \cite{Textures-SAR_2021}.
Next, we discuss a few specific examples of application of these concepts in different fields.

\section{Uncovering similarities between neurons and lasers}

While neurons and lasers are, at first sight, remarkably different dynamical systems, ordinal analysis has allowed to uncover some unexpected similarities~\cite{njp_2019}. By analyzing the statistics of sequences of spikes, simulated with neuronal models and recorded experimentally, in the intensity emitted by a semiconductor laser with feedback, {{conditions have been identified}} in which the sequences of neuronal and optical spikes have the same hierarchy of ordinal probabilities, i.e., the same order in the probabilities, that reveal the same more-expressed and less-expressed patterns in the spike sequences (see Fig. 4a in~\cite{srep_2014} and Fig. 3a in~\cite{pre_2016}). 

In both spike trains (recorded from the laser output and simulated with a neuronal model), when using OPs of length $D=3$ (that give six possible OPs, which we may label as $123$, $132$, $213$, $231$, $312$ and $321$) {{ it has been found}} that four out of the six patterns form two ``clusters'' with very similar probabilities: $P(132)\sim P(213)$ and $P(312)\sim P(231)$. This property holds in other experimental time series~\cite{pra_2018} and in simulations of a minimal model represented by a ``circle map''~\cite{srep_2014}. 

Since neurons encode and transmit information in sequences of spikes, uncovering similarities between neuronal spike trains and optical spike trains may open a path for implementing optical neurons, that genuinely mimic neuronal responses and are able to use neuronal mechanisms to process information, as efficiently as biological neurons, but orders of magnitude faster. Progress in this direction requires a good understanding of the neuronal encoding mechanisms and how they can be implemented in laser systems. 

Using neuronal models {{it has been shown}} that, when a weak (subthreshold) periodic input is perceived by a stochastic neuron, the neuron fires a sequence of spikes with specific properties of ordinal probabilities. The ordinal probabilities depend on the amplitude and on the period of the input signal~\cite{pre_2016, chaos_2020} and thus, they may be informative of these features of the signal. {{It was}} also found that this encoding mechanism can be enhanced in an ensemble of neurons, when they all perceive the weak signal~\cite{cnsns_2020}. Thus, a neuronal ensemble could encode, in time-varying ordinal probabilities, time-varying features of aperiodic signals. 

Exploiting this coding mechanism to implement laser systems that use neural information coding requires a careful comparison of the properties of the ordinal probabilities computed from experimental optical spike sequences and those simulated neuronal ones, that are emitted in response to different types of aperiodic signals. While the two distributions of ordinal probabilities, $P_l$ and $P_n$, computed from the output of the laser and from the output of a neuron model might have similar values of permutation entropy and statistical complexity, there may be some differences that are not captured by these quantifiers. Therefore, additional research is needed in order to better understand how to detect and quantify similarities, and also significant differences, in the values of the ordinal probabilities.

\section{Some biomedical applications}

Entropy plays an essential role in biology. While at first this seems at odd with its concept of being randomness and disorder, living entities are constantly swimming against the inexorable increase in entropy mandated by thermodynamics. In other words, being alive implies maintaining control in a sea of disorder; hence, the analysis of how (and how much) such control is enforced gives hints about how healthy the organism is. Permutation entropy further allows to add a new dimension to the analysis: the temporal one, i.e. how the system copes with, or depends on, its history.

Analysis of biomedical time series started soon after this methodology was proposed, to the best of our knowledge with the study of electroencephalography data near epileptic events \cite{keller2004distances}. This was closely followed by a plethora of applications, spanning from heart rate \cite{frank2006permutation}, gene expression \cite{sun2010complexity}, electromyographic (muscle response) data \cite{sanjari2010local}, intracranial pressure \cite{kalpakis2012outcome}, to human gait \cite{zanin2018characterizing}. Many success stories have emerged, as well as many tools to diagnose a wide range of pathologies \cite{adjei2015intracranial, liu2017multiscale, cuesta2018classification, cseker2021complexity}.

Medicine is a fast evolving field, and it will require a parallel evolution in how we analyse biomedical data. We foresee two forces that may change how permutation patterns and associated metrics will be used in this context. 

The first one is exogenous, and relates to how we can monitor the human body. Novel technologies are allowing moving from a lab-centric approach, to a scenario in which smart devices (as e.g. smart watches and activity monitors) provide non-invasive ways of monitoring patients throughout their daily life \cite{axisa2005flexible}. This implies, on one hand, a huge data quality improvement: long time series can, for the first time, be recorded in a natural environment. On the other hand, and almost contradictory, this comes at the cost of a reduction in quality: sensors of this kind can be noisy and may generate artefacts or missing data. Little is known about how these aspects can affect the estimation of permutation patterns and associated metrics \cite{mccullough2016counting, sakellariou2016counting, kulp2016using}, especially in a context of non-stationarity. Different sensors can also provide information with different time resolutions, both because of their technical specifications and of the time scales that are characteristic of the processes they measure. To illustrate, postural data can be registered several times per second \cite{zanin2018characterizing}, while glucose levels can be sampled every five minutes \cite{cuesta2018classification}. While permutation patterns can be assessed on multivariate time series \cite{he2016multivariate, schlemmer2018spatiotemporal}, how such heterogeneity can be dealt with is yet not clear. 

This shift in data may also be accompanied by an endogenous change in the way permutation patterns are studied. Specifically, such patterns constitute the building blocks for many complementary metrics: from simple entropies, complexity and irreversibility \cite{zanin2018assessing, martinez2018detection} metrics, up to the reconstruction of transition networks \cite{pessa2019characterizing, olivares2020contrasting}. With the notable exception of the first two \cite{Complex_PRL_2007}, little is known about how these metrics interact; future research works can then be focused on understanding how complementary the information they provide is, up to the possibility of constructing an entropy-irreversibility plane.

\section{Application to complex networks and neuroscience}

A relevant research field where ordinal methods could be expanded is the study of ensembles of coupled dynamics systems, particularly those with nontrivial topologies as complex networks. For small groups of elements, several studies have implemented new methods to detect correlations between the temporal series of coupled dynamical systems. The ordinal techniques are generally faster and computationally cheaper than standard correlation methods in those cases. Bahraminasab et. al. \cite{Bahraminasab2008} developed the {\it permutation information} approach to quantify the directionality of coupling between interacting oscillators. In. Ref. \cite{Montani2015} the authors use a multi-scale version of this approach for discriminating the time delay between two coupled time series, applying the method to reveal delayed and anticipated synchronisation between the dynamics of brain areas. The coherence level can be also measured in terms of the {\it ordinal synchronisation}  \cite{Echegoyen2019}, with the advantage over non-ordinal methods of being able to detect signatures of both phase and antiphase synchronisation in weakly coupled systems in a very fast way. Another possible route is using {\it multivariate ordinal patterns} to analyse several time series simultaneously. This method has allowed detecting the proximity to synchronisation in small groups of chaotic coupled nodes \cite{Zhang2017}, and should be extensible to large ensembles. 
 
However, the use of ordinal methods in studying large complex networks dynamics has been limited so far, but it is a promising tool for the important issue of inferring the unknown underlying structure of connections. In some cases, the {\it ordinal mutual information} allows for the complete inference of the underlying network \cite{Rubido2014,Tirabassi2015}, provided an appropriated choice of the observable variable and dynamical range. 

Beyond these methods based on pairwise interactions, every single element of the ensemble is encoding in its dynamics the interaction with its environment, deviating from the isolated node dynamics in a node-specific way. The fact is exploited in Ref. \cite{Tlaie2019dy,Letellier2021} to obtain information about the network structure. The principal finding is that in the weakly coupled regime, the topological centrality of each node can be inversely correlated to its statistical complexity, therefore acting as a straightforward proxy for the degree distribution. This correlation appears very general for a wide range of dynamics and network structures, and it allows to detect the network hubs without references to site-to-site correlations. The method is applied in Ref. \cite{Tlaie2019st} to analyse the structure of networks of {\it in vitro} cultured neurons from invertebrates, showing that the structural information can be retrieved even in stochastic environments.    

The robustness of ordinal methods to noise and their ability to discriminate between dynamical states make them particularly well suited to tackle the analysis of brain signals in the framework of complex networks. The interest of the neuroscientist for the ordinal methods is continuously spreading to face new problems as early cognitive diseases \cite{Echegoyen2020}, analysis of resting states\cite{Quintero-Quiroz2018,MEG_Chaos_2020}, cognitive reserve \cite{Martinez2018}, stress \cite{Garcia-Martinez2017}, or epilepsy \cite{Yao2020}, to mention some examples. Very recently, the newest techniques such as ordinal networks and multivariate ordinal analysis  \cite{Varley2021a, Varley2021b, chavez2020ordinal} have been incorporated into the tool collection of neuroscience, and it is to be expected that they become fundamental in the next decade. 

\section{Applications to sport science}

During the last years, sports have been benefited from new technologies that extract priceless information about what happens during both the competition and training \cite{gudmundsson2017}.
The access to this data has awakened the interest of different scientific areas in sports \cite{hodson2021,maguire2011,wu2020,da2013}, and nonlinear dynamics and statistical physics are not an exception.
A new point of view arises when considering a collective game as a physical system composed of players/athletes interacting under specific rules and evolving in time in a constrained space \cite{buldu2018,buldu2021}.
Under this scope is not strange to use network science and time-series analysis to advance in the understanding of the team's organization, performance, scoring processes, or the evaluation players' physical condition. \cite{herrera2020,duch2010,lames2006,downward2016,witte2001}.
For example, information theory has been widely used to tackle the nonlinearity aspects of teams' performance \cite{borrie2002,jonsson2010,camerino2012,borooah2012,silva2016,martins2020,ruth2020,chacoma2021,pereira2021}.
However, the scientific literature contains few studies assessing the information content of teams by using ordinal patterns~\cite{martinez2020}. Specifically, in football, OPs have been proposed as a way to identify what variables of a team should be observed in order to understand team performance. In this way, {{this}} recent paper has analysed the Spanish football league to describe the dynamics of players' interactions~\cite{martinez2020}. In this work, the authors construct passing networks whose structure evolves during a match and map the evolution of their features, which are endowed with the team's structural progression, thus capturing its information content.
Results show both a negative correlation of networks' features dynamics in the entropy-complexity plane; and how the entropy of centrality measures (center of mass) of teams is positively (negatively) influenced by their average number of passes.

Nowadays, the door is open for further studies using OPs methods for the benefit of sports science. To name but a few: new studies might evaluate whether the inner dynamics of teams would be influenced by the context of the game, e.g., the score, the players on the pitch, the ranking of the teams, etc. At the same time, OPs methods could be used to analyze the evolution of different game events, such as faults, tackles, shots, or goals.  
It would be also interesting to observe the change of entropy and complexity of teams dynamics when playing at different moments of the season, or different scenarios (e.g., Champions league, vs country's national league), altogether could be framed in the causal  $(H \times C)$ plane. One could even think of using information theory to compare different collective games and investigate, for example, how the effective position of teams evolves in time and space and its relation with the pitch/court size \cite{martinez2020,chacoma2021}.
The use of ordinal networks \cite{herrera2020symbolic,masoller2015quantifying} or measures of statistical irreversibility \cite{martinez2018detection,zanin2021algorithmic} could also offer new ways of studying the inner dynamics of games with this kind of data.

Furthermore, this kind of analysis could also be used to unveil the synchronization between teams. The use of information-based correlations of OPs \cite{cazelles2004,Echegoyen2019} might lead to interesting results when observing the degree of cohesion between fluctuations of a network parameter of two teams in a match.

From the spatial point of view, there is debate around players' heatmaps obtained from events or tracking datasets \cite{buldu2020}. The former with spatio-temporal knowledge of players' actions (passes, faults, goals); and the latter with the localisation of players and with a resolution of a few centimeters and 25 frames per second \cite{garrido2021}. Heatmaps characterise the spatial role of players by mapping their activity into matrices of pixels. These kinds of matrices have previously been analysed by a bidimensional adaptation of OPs in other contexts \cite{ribeiro2012,sigaki2018,pessa2021ordpy}. Nonetheless, the use of planar patterns in heatmaps would shed new insights around the event-vs-tracking data debate to hopefully understand and detect the team's formations from a nonlinear perspective.

It is also well-known that some teams use to have a particular game style. Some of them prefer long-range passes with large time lapses between them, and others adopt a composite style, e.g., ball possession mainly driven by long-playing sequences. In a mimic of biological systems, the act of passing the ball can be viewed as a neuronal spike solely activated when the neuron (player) conveys information (pass) to another one. These all-or-none ``signals" provide information of the neural culture (the team, in this analogy) by studying the interspike intervals, i.e., the interludes between two successive events \cite{sabatier2004}. The information content of the time series of inter-passes intervals could shed light on the game style of a team by characterising the entropy and complexity of OPs of these passing sequences.

In view of all, the application of OPs to sports science will offer a complementary perspective regarding team or individual performance in the years to come, and it can be considered as an emergent branch with potential and interesting results to unveil the nonlinear dynamics hidden in collective sports.

\section{Outlook}

We end this perspective article by proposing some, in our opinion, promising research lines.
\begin{enumerate}
\item Over the years, extensions of the BP approach have been proposed (e.g., ~\cite{fadlallah2013weighted,Small_2014, Zunino_2015, Politi2017}). Their advantages and drawbacks, in particular in terms of their computational requirements and robustness to noise are not yet fully understood.

\item New extensions to multivariate signals and ensemble dynamics, and
new applications of the causal complexity -- permutation entropy plane. For example, for tracking changes in the teams' dynamics of collective sports at different seasons or different scenarios. It could also be used to compare reached levels of these quantifiers extracted from planar OPs applied on data of events, vs tracking of players.  

\item A better understanding of how to deal with missing data or equal values, non-equi-sampled values, etc.

\item Alternative approaches for defining ordinal pattern networks (which pattern follows each pattern) that best capture the correlational structures in the data. 

\item A better understanding of how to assign confidence intervals to the ordinal probabilities and to the information-theory measures (causal complexity, permutation entropy, Fisher entropy, etc.). The work done by Frery group~\cite{Frery_2022}, for uncorrelated noises opens a promising path in this direction.

\item {{More research is needed to better understand the statistical properties of the ordinal patterns, and in particular, the values of the ordinal probabilities. If the existence of ordinal patterns with similar probabilities turns out to be a generic property of certain dynamical systems, this should be taken into account when combining ordinal analysis with machine learning, as a proper selection of the input features (ordinal probabilities) will avoid feeding the machine learning algorithm with redundant features.}}
\end{enumerate}




\acknowledgments
The authors acknowledge hospitality of the Max Planck Institute for the Physics of Complex Systems, where a collaboration was established and this work started. C.M. acknowledges funding from the ICREA ACADEMIA program, Generalitat de Catalunya; M.Z. from the European Research Council (ERC) under grant agreement No 851255; M.Z. and J.M. from the Spanish State Research Agency through Grant MDM-2017-0711 funded by MCIN/AEI/10.13039/501100011033; I.L. from the Spanish State Research Agency through Grant PID2020-113737GB-I00. J.M acknowledges J. M. Buld\'u for valuable comments and J. J. Ramasco for his assistance.

\bibliographystyle{eplbib}
\end{document}